\documentclass[final,3p,times]{elsarticle}

\usepackage{amssymb}
\usepackage{amsmath}
\usepackage{graphicx}
\usepackage{xcolor}
\usepackage{placeins}
\usepackage{lineno}        

\journal{Journal of Subatomic Particles and Cosmology}

\newcommand{\sNN}{\sqrt{s_{\rm NN}}}

\begin{document}

\begin{frontmatter}

\title{Observation of Strong Directed Flow for $\phi$ Mesons in the High Baryon Density Region at RHIC}

\author[aaa]{Guangyu Zheng (for the STAR Collaboration)}
\affiliation[aaa]{organization={School of Nuclear Science and Technology, University of Chinese Academy of Sciences},
             addressline={Beijing 100049},
             country={P.R. China}}

\begin{abstract}
We report measurements of the rapidity-odd directed flow ($v_{1}$) of $\phi$ mesons in 10--40\% Au+Au collisions at $\sNN=3.0$--4.5 GeV from the STAR fixed-target program at RHIC, together with the collider-energy measurement at $\sNN=7.7$ GeV. Despite its mesonic nature, the $\phi$-meson $v_{1}$ exhibits a magnitude comparable to that of protons and $\Lambda$ baryons and is significantly larger than that of $K^{0}_{S}$ mesons in the high-baryon-density region. Comparisons with UrQMD calculations indicate that the observed $\phi$-meson directed flow is consistent with production through high-mass baryon resonances and the collective motion of baryons influenced by baryonic mean-field interactions. These results provide new experimental constraints on baryon-resonance production and strange-hadron transport dynamics in dense QCD matter.
\end{abstract}


\end{frontmatter}


\section{Introduction}

Relativistic heavy-ion collisions probe strongly interacting matter at different temperatures and baryon chemical potentials. Directed flow, $v_{1}$, is the first harmonic of the particle azimuthal distribution relative to the reaction plane. Generated mainly by the early sideward deflection of compressed matter, it is sensitive to the dynamics and equation of state of baryon-rich matter~\cite{Stoecker1986,Brachmann2000,Bzdak2020}. STAR observed a nonmonotonic collision-energy dependence of proton and $\Lambda$ $dv_{1}/dy$ in the RHIC Beam Energy Scan, motivating further studies of baryonic interactions and mean-field effects at high baryon density~\cite{STARv1BES2014,STARv1BES2018,STARFXT}.

The $\phi$ meson is a distinctive probe as the lightest hidden-strangeness ($s\bar{s}$) vector meson, with a mass close to that of the proton. Its relatively long lifetime ($\sim 46$ fm/$c$) and weak effective interaction with nucleons make it less sensitive to late-stage hadronic rescattering than lighter mesons~\cite{PDG2024,Wood2010}. Near threshold, transport studies indicate that $\phi$ mesons can be produced predominantly through high-mass baryon-resonance decays~\cite{Steinheimer2016,Steinheimer2025}, potentially linking their directed flow to the collective motion of baryon-rich matter. Thus, low-energy $\phi$-meson $v_{1}$ measurements probe baryon-resonance dynamics and strange-hadron transport.

These proceedings present rapidity-odd $\phi$-meson $v_{1}$ in mid-central Au+Au collisions at $\sNN=3.0$, 3.2, 3.5, 3.9, and 4.5 GeV. Results are compared with protons, $\Lambda$ baryons, $K^{0}_{S}$ mesons, and UrQMD calculations. Further analysis details are given in Ref.~\cite{STAR:2026ley}.

\section{Experiment and analysis}

The data were collected during the 2019--2021 STAR fixed-target Beam Energy Scan, with an Au beam incident on a stationary Au target near the Time Projection Chamber (TPC) entrance~\cite{Anderson2003}. This analysis uses 10--40\% central Au+Au collisions. Charged kaons are identified using TPC specific ionization and barrel/endcap time-of-flight measurements~\cite{Llope2012}.

The $\phi$ meson is reconstructed via $\phi\rightarrow K^{+}K^{-}$. Pseudorapidity $\eta$ is defined in the laboratory frame relative to the displaced fixed-target vertex, while rapidity $y$ is defined in the center-of-mass frame, positive along the incident Au beam direction. The first-order event plane $\Psi_{1}$ is reconstructed with the east EPD~\cite{AdamsEPD2020}, whose effective fixed-target acceptance is $-5.3<\eta<-2.6$. Directed flow is evaluated as $v_{1}=\langle\cos(\varphi-\Psi_{1})/R_{1}\rangle$, where $\varphi$ is the particle azimuth and $R_{1}$ is the event-plane resolution. The event-plane orientation follows the projectile-spectator deflection, giving $v_{1}<0$ at negative $y$ for a positive rapidity-odd slope~\cite{STARFXT}. The resolution (55--75\%) is determined using three subevents (two EPD and one TPC)~\cite{Poskanzer1998}.

\begin{figure}[!tbp]
\centering
\includegraphics[width=0.75\linewidth]{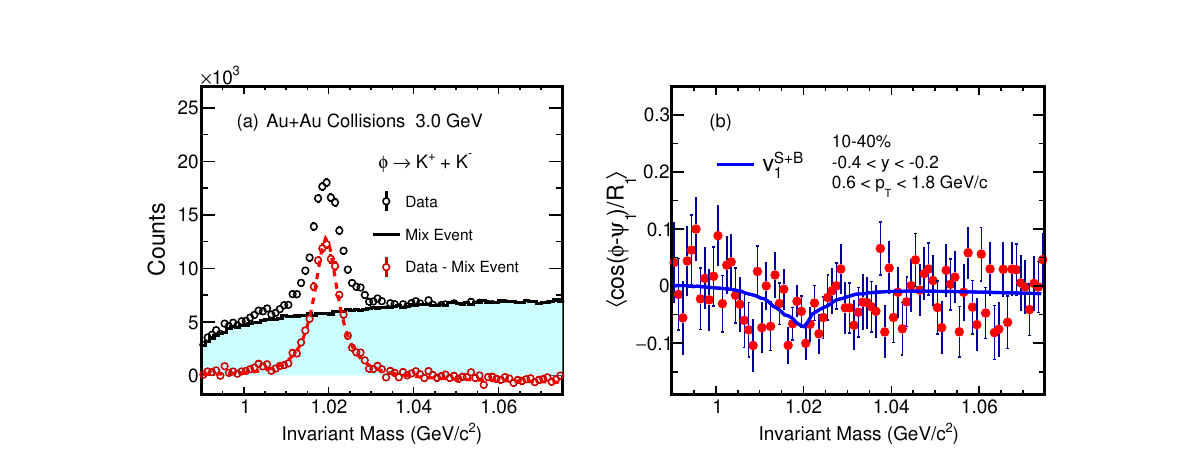}
\caption{(a) Invariant-mass distribution of $K^{+}K^{-}$ pairs used for $\phi$-meson reconstruction in $\sNN=3.0$ GeV Au+Au collisions (10--40\% centrality). (b) Directed flow $v_{1}$ as a function of invariant mass in the same kinematic region. The fit separates the signal and background contributions to $v_{1}$ and validates the extraction of the $\phi$-meson directed flow.}
\label{fig:invmass}
\end{figure}

Combinatorial backgrounds are estimated using the mixed-event technique, which combines $K^{+}$ and $K^{-}$ candidates from different events with similar centrality and event-plane angle to construct the uncorrelated background~\cite{Drijard1984}. The $\phi$-meson $v_{1}$ is extracted with the invariant-mass method~\cite{Borghini2004},
\begin{equation}
  v_{1}^{S+B}(M) =
  v_{1}^{S}\frac{S}{S+B}(M)
  +v_{1}^{B}(M)\frac{B}{S+B}(M),
  \label{eq:invmass}
\end{equation}
where $M$ is the $K^{+}K^{-}$ invariant mass, $S$ and $B$ are the signal and background yields, and $v_{1}^{S}$ and $v_{1}^{B}$ are their respective directed-flow components. The signal fraction $S/(S+B)$ is determined from the invariant-mass analysis using a Breit--Wigner signal shape and a linear residual background after mixed-event subtraction. The background flow $v_{1}^{B}(M)$ is parameterized as a first-order polynomial in $M$ and fitted together with $v_{1}^{S}$ to the measured $v_{1}^{S+B}(M)$. Fig.~\ref{fig:invmass} illustrates the procedure at $\sNN=3.0$ GeV. The $p_{T}$-integrated $v_{1}$ is measured for $0.6<p_{T}<1.8$ GeV/$c$ and $-1<y<0$, with efficiency corrections in $p_T$ and rapidity. The midrapidity slope is obtained from a linear fit through the origin. Systematic uncertainties are evaluated by varying track-quality cuts, particle identification, invariant-mass extraction, and event-plane resolution correction.

\section{Results and discussion}

\begin{figure}[!tbp]
\centering
\includegraphics[width=0.8\linewidth]{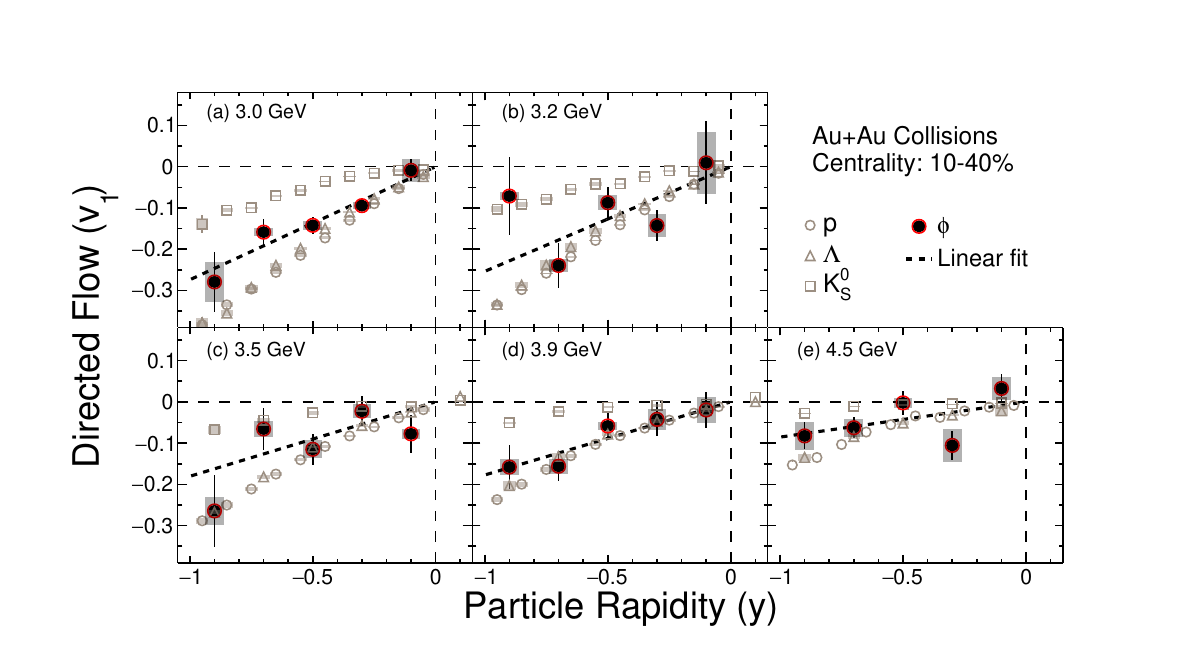}
\caption{Directed flow $v_{1}$ as a function of rapidity $y$ for $\phi$ mesons, protons ($p$), $\Lambda$ baryons, and $K^{0}_{S}$ mesons in 10--40\% Au+Au collisions at $\sNN=3.0$--4.5 GeV. The black dashed lines represent first-order polynomial fits used to extract the $\phi$-meson midrapidity slope $dv_{1}/dy$. Statistical and systematic uncertainties are indicated by bars and shaded bands, respectively.}
\label{fig:v1y}
\end{figure}

Fig.~\ref{fig:v1y} shows the $\phi$-meson $v_{1}(y)$ together with proton, $\Lambda$, and $K^{0}_{S}$ results in 10--40\% Au+Au collisions at $\sNN=3.0$--4.5 GeV. The $\phi$ meson exhibits a positive midrapidity $v_{1}$ slope at all five energies, with a magnitude comparable to those of protons and $\Lambda$ baryons but significantly larger than that of $K^{0}_{S}$ mesons. The $v_{1}$ magnitude generally decreases with increasing collision energy.

\begin{figure}[!tbp]
\centering
\includegraphics[width=0.5\linewidth]{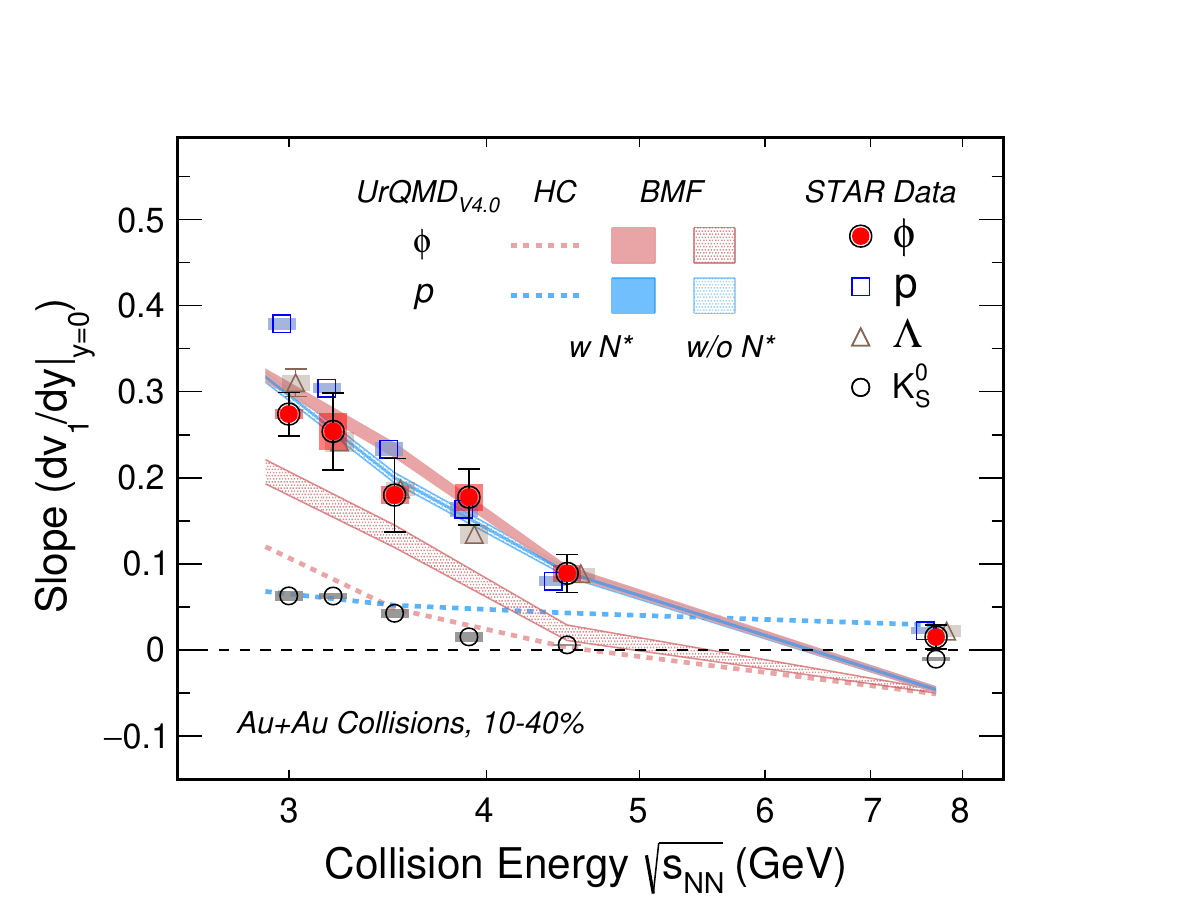}
\caption{Collision-energy dependence of the midrapidity $v_{1}$ slope $dv_{1}/dy|_{y=0}$ for $\phi$ mesons, $K^{0}_{S}$ mesons, protons, and $\Lambda$ baryons in 10--40\% Au+Au collisions. Statistical and systematic uncertainties are indicated by bars and shaded bands, respectively. The curves show UrQMD v4.0 calculations in the hadronic-cascade mode (HC) and with baryonic mean-field interactions (BMF). For $\phi$ mesons, the BMF calculation is shown both with and without high-mass $N^{*}$ contributions.}
\label{fig:energy}
\end{figure}

Fig.~\ref{fig:energy} compares the midrapidity $dv_{1}/dy$ slopes with UrQMD v4.0 calculations~\cite{Steinheimer2016,Steinheimer2025}. At the lowest energies, the $\phi$-meson slope exceeds that of $K^{0}_{S}$ by a factor of three or more. The baryonic mean-field (BMF) treatment uses momentum-dependent potentials based on a chiral mean-field (CMF) model~\cite{SteinheimerCMF2025}. Within UrQMD, near threshold $\phi$ production proceeds through high-mass $N^{*}$ resonance decays, with production and inverse absorption channels constrained by hadronic cross sections and detailed balance~\cite{Steinheimer2016,Steinheimer2025}. The hadron-cascade mode without BMF underestimates the $\phi$-meson and proton slopes, while including BMF and high-mass resonances reasonably describes their energy dependence at $\sNN=3.0$--4.5 GeV. At 7.7 GeV, both configurations underestimate the measured $\phi$-meson slope.

To examine the resonance contribution, the $N^{*}\rightarrow N\phi$ channel is suppressed while retaining BMF. The proton $v_{1}$ slope remains nearly unchanged, whereas the calculated $\phi$-meson $v_{1}$ is strongly reduced. This indicates the importance of this channel within UrQMD, without uniquely establishing the underlying production mechanism. A possible production path is
\begin{equation}
 N+N\rightarrow N^{*}+N\rightarrow N+\phi+N.
\label{eq:nstar}
\end{equation}
High-mass resonances, such as $N^{*}(1990)$ and $N^{*}(2250)$, may transfer collective baryonic motion to the emitted $\phi$ mesons. This interpretation remains subject to uncertainties in the mean-field parametrization, resonance-production and decay inputs, and $\phi$--nucleon interactions. The scenario is also consistent with the enhanced $N(\phi)/N(K^{-})$ and $N(\phi)/N(\Xi^{-})$ yield ratios observed by STAR at $\sNN=3$ GeV~\cite{STARphiYield}.

\FloatBarrier

\section{Summary}

We present the STAR measurement of the rapidity-odd directed flow of $\phi$ mesons in 10--40\% Au+Au collisions at $\sNN=3.0$--7.7 GeV. The $\phi$-meson midrapidity $v_{1}$ slope is positive and comparable to those of protons and $\Lambda$ baryons and significantly larger than that of $K^{0}_{S}$ mesons at the lowest energies. Comparisons with UrQMD calculations are consistent with a scenario in which high-mass baryon resonances and collective baryonic motion influenced by mean-field interactions contribute to the observed $\phi$-meson directed flow; the model comparison does not uniquely identify the production mechanism. The $\phi$ meson therefore provides a unique probe of baryon-resonance dynamics and strange-hadron transport in the highest-baryon-density region accessible at RHIC.

\section{Acknowledgments}

This work is partially supported by the National Natural Science Foundation of China under Grant No.~12305146 and the Project for Young Scientists in Basic Research (No.~YSBR-088) of the Chinese Academy of Sciences.

\end{document}